# Thermodynamic Analysis of Titanium Removal from Molten iron Smelted with Large Amounts of Sodium Additives

Zhiwei Bian [a, b, c], Desheng Chen [a, b, d, *], Lina Wang [a, b, d], Hongxin Zhao [a, b, d], Yulan Zhen [a, b, d], Tao Qi [a, b]

[a] *National Engineering Laboratory for Hydrometallurgical Cleaner Production Technology, Institute of Process Engineering, Chinese Academy of Sciences, Beijing 100190, China*

[b] *Key Laboratory of Green Process and Engineering, Institute of Process Engineering, Chinese Academy of Sciences, Beijing 100190, China*

[c] *University of Chinese Academy of Sciences, Beijing 100190, China*

[d] *HeBei ZhongKe TongChuang Vanadium Titaninum Science and Technology co., LtD., Hengshui 131100, Hebei, China*

**Abstract**

High purity iron is obtained from vanadium-titanium magnetite (VTM) by one-step coal-based direct reduction-smelting process with coal as reductant and sodium carbonate ($Na_2CO_3$) as additives. Industrial experiments show that the process of treating molten iron with a large amount of $Na_2CO_3$ is effective in removing titanium from molten iron. However, the studies are rarely conducted in thermodynamic relationship between titanium and other components of molten iron, under the condition of a large amount of $Na_2CO_3$ additives. In this study, through the thermodynamic database software Factsage8.0, the effects of melting temperature, sodium content and oxygen content on the removal of titanium from molten iron are studied. The results of thermodynamic calculation show that the removal of titanium from molten iron needs to be under the condition of oxidation, and the temperature should be below the critical temperature of titanium removal (the highest temperature at which titanium can be removed). Relatively low temperature and high oxygen content contribute to the removal of titanium from molten iron. The high oxygen content is conducive to the simultaneous removal of titanium and phosphorus from molten iron. In addition, from a thermodynamic point of view, excessive sodium addition inhibits the removal of titanium from molten iron.

**Introduction**

VTM is rich in iron, vanadium, titanium and other high-value elements, which has high comprehensive utilization value, and is widely distributed in the world with large reserves [1]. In China, VTM accounts for more than 60% of vanadium resources and 90% of titanium resources [2]. At present, the main utilization methods of VTM are blast furnace (BF) process and direct reduction (DR) process. In the BF process, most of iron and vanadium in VTM enter the molten iron to produce pig iron containing vanadium, while titanium enters slag to form titanium slag, which contains about 22%-25% titanium dioxide ($TiO_2$) [3]. The distribution of titanium components in titanium slag is extremely diffuse and the mineral interface combination is very complex, which makes it difficult to recover the titanium components from titanium slag [1, 5~6]. In Panzhihua area of Sichuan Province, China, VTM is mainly used in BF process, which produces a large amount of titanium slag every year [4]. For the DR process, it is still in the development and research stage, the technology is not very mature, and the titanium slag produced is difficult to use. Therefore, large amounts of accumulated titanium slag caused serious waste of titanium resources and pollution of environment.

Due to the performance requirements of high-end steels, high-end steels shall contain low-percentage of titanium content ($\leq$ 0.0008%), which means that they strictly require very low-percentage titanium content in raw material - molten iron. Ti-bearing inclusions, such as titanium nitride (TiN) and titanium carbonitride ($TiC_yN_{1-y}$), that exist in high-end steel are harmful to fatigue properties and drawing performance of steels [7~10]. The larger the titanium inclusion size is, the more harmful it is to the steel property [10]. Some researches indicate that these Ti-bearing inclusions are mainly formed during solidification of molten steel, and with the decrease of titanium content, the size and quantity of Ti-bearing inclusions will decrease [8, 10~12]. Therefore, to reduce the titanium content in molten iron as much as possible is an important measure to produce high-performance steel.

In order to efficiently utilize the VTM in China, a lot of research works has been done in our laboratory [13~18]. A new metallurgical process to recover vanadium and titanium, and to produce high purity iron simultaneously from VTM has been developed [13]. The process of high purity iron preparation of the new technology is shown in Fig. 1.

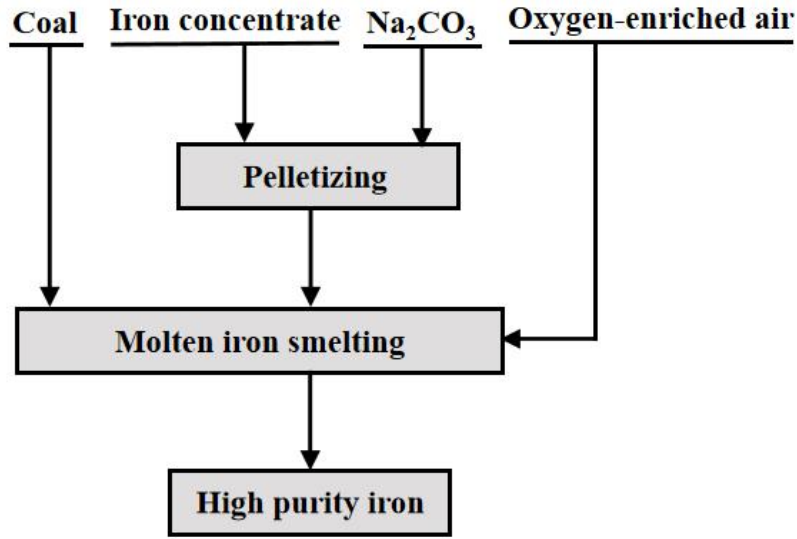

**Fig. 1.** Flow chart of the new process of high purity iron preparation.

In this process, the titanomagnetite concentrates are mixed with large amounts of $Na_2CO_3$ additives to make pellets firstly and the mass ratio of concentrates and $Na_2CO_3$ is nearly 1:1. The pellets and coal are then mixed into a melting furnace, where they are melted with oxygen-rich air at about 1500K. The mass ratio of concentrates and coal is about 1: 0.3. Then, the high purity iron with low sulfur, phosphorus, and titanium can be obtained after casting. Although there are a lot of researches in the sodium treatment of molten iron and it is proved that sodium salt has a good effect of impurity removal, the amount of sodium salt added in these studies is very small (＜10%) [19~24]. In our researches, such large amounts of $Na_2CO_3$ additives can reduce the titanium, sulfur, phosphorus content in molten iron to very low. However, the thermodynamic effect of removing titanium from molten iron by a large amount of $Na_2CO_3$ additives is not very clear.

In this work, the effects of temperature, oxygen content and sodium content on titanium removal are studied, and the relationship between oxygen content and the critical temperature of titanium removal from molten iron is determined by thermodynamic database software Factsage8.0. In addition, the oxidation conditions of simultaneous detitanationn and dephosphorizatio of molten iron treated by large amounts of sodium are also studied. The objective of this work is to provide a theoretical basis for the production process of high purity iron with large amounts of $Na_2CO_3$ as additives.

**Thermodynamic calculation**

*Chemical compositions of molten iron in thermodynamic calculation*

The new process is successfully carried out in a pilot trial and the high purity iron of ultra-low sulfur, phosphor, and titanium is successfully obtained. Typical main chemical compositions of molten iron in the new process are selected for analysis, as shown in Table 1.

**Table 1**

Chemical compositions of the molten iron, wt%.

| Elements | C | V | Mn | Cr | Ti | Si | P | S |
|---|---|---|---|---|---|---|---|---|
| Content | 4.11 | 0.36 | 0.16 | 0.076 | 0.013 | 0.011 | 0.001 | 0.0005 |

*Software settings in thermodynamic calculation*

Equilib module of Factage8.0 was used for calculation [25]. First of all, the content of each element is set according to the chemical compositions of the molten iron as shown in Table 1. Oxygen and sodium contents are set according to the actual production data to clearly identify the effect of changes in oxygen and sodium contents on titanium removal from the molten iron. The specific database and temperature and other conditions are set as follows:

Database: FTmisc; FToxid; FactPS;

Temperature: 1400K; 1450 K; 1500 K; 1550 K; 1600 K; 1650 K;

Solution phase setting: FToxid-SLAGA; FTmisc-FeLQ;

Compound setting: pure solids;

Equilibrium: normal;

Weight of molten iron: 100g.

**Results and discussions**

*Effect of temperature on titanium removal*

In order to accurately evaluate the effect of temperature on titanium removal from molten iron during sodium treatment process, the addition of sodium in molten iron is set at 40g. The oxygen content in molten iron is set at 1g, and the content of other components is set according to Table 1. Fig. 2 shows the effect of temperature on titanium removal from molten iron.

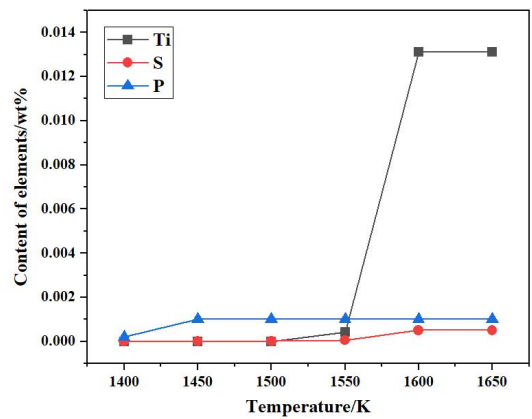

**Fig. 2.** Effect of temperature on titanium removal from molten iron.

As can be seen from Fig. 2, when the sodium content and oxygen content are constant, the sulfur and phosphorus content in molten iron do not change significantly when the temperature rises from 1400K to 1650K. When the temperature is lower than 1550K, the change of titanium content in molten iron is small. However, when the temperature is higher than 1550K, the titanium content in molten iron rises rapidly and remains stable after the temperature reaches 1600K. In this process, 1550K can be considered as a critical temperature for titanium removal. The removal of titanium from molten iron needs to be below the critical temperature. This means that relatively low temperature is favorable for titanium removal from molten iron. This result is consistent with other findings in literatures [26~27]. Therefore, in the process of sodium treatment of molten iron, relatively low temperature operation should be used appropriately, so as to facilitate the removal of titanium from molten iron.

Similarly, the above critical temperature of titanium removal exists under other conditions of sodium and oxygen content. When the sodium content is from 40g to 60g, and the oxygen content is from 0.1g to 2g, the effect of temperature on the sulfur, phosphorus and titanium in the molten iron components is similar, and there is a critical detitanation temperature. When the temperature of molten iron is lower than the critical temperature of titanium removal, the content of titanium in molten iron is relatively low. However, as long as the temperature of molten iron is higher than the critical detitanation temperature, the content of titanium in molten iron will increase rapidly.

*Effect of oxygen content on titanium removal*

To evaluate the effect of oxygen content on titanium removal from molten iron, the sodium content of molten iron is fixed at 50g. The oxygen content in molten iron was set at 0.1g, 0.5g, 1g, 1.5g and 2g respectively. Fig. 3 shows the effect of oxygen content on the critical temperature of titanium removal from molten iron.

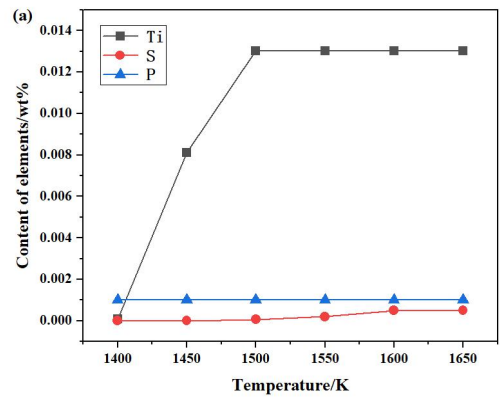

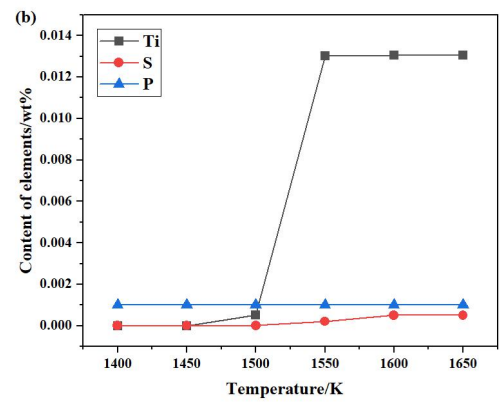

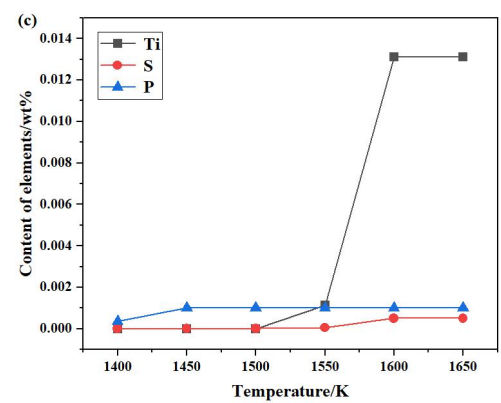

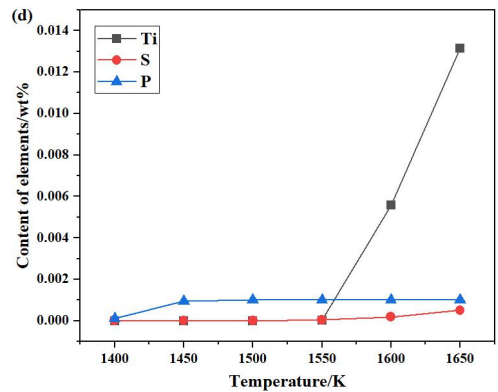

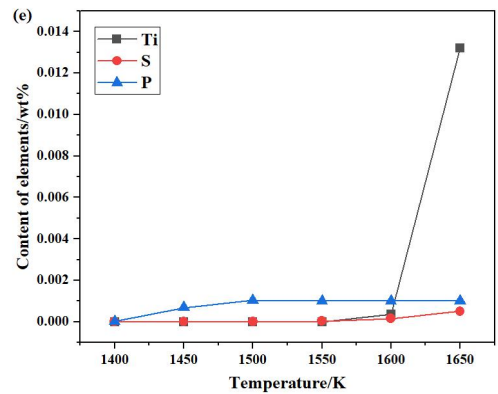

**Fig. 3.** Effect of oxygen content on the critical temperature of titanium removal from molten iron, the oxygen content at (a) 0.1g; (b) 0.5g; (c) 1g; (d) 1.5g; (e) 2g.

It can be clearly seen from Fig. 3 that with the increase of oxygen content in molten iron, the critical detitanation temperature increases. As the oxygen content in molten iron increases from 0.1g to 2g, the critical detitanation temperature increases from 1400K to 1600K.

In order to more clearly demonstrate the influence of oxygen content on titanium removal from molten iron, the amount of sodium added in molten iron is set at 50g, and the temperature is set at 1550K. Fig. 4 shows the effect of oxygen content on titanium removal from molten iron under the conditions of sodium content of 50g and 1550K. As can be seen Fig. 4, when the oxygen content is higher than 0.4g, the titanium content in molten iron begins to decline. At the oxygen content of 1g, most of the titanium in the molten iron has been removed. When the oxygen content is higher than 1g, the titanium in molten iron is further removed to extremely low with the increase of oxygen content. This indicates that due to the oxidation reaction between oxygen and titanium, high oxygen content contributes to the removal of titanium from molten iron.

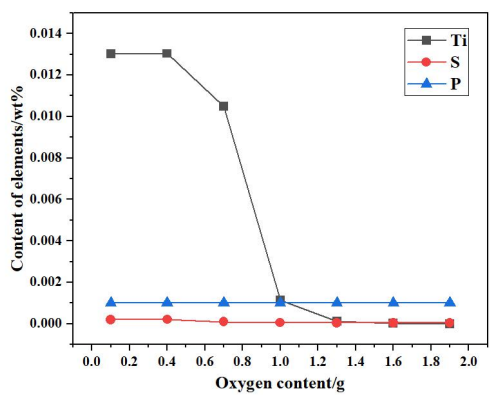

**Fig. 4.** Effect of oxygen content on titanium removal from molten iron at the temperature of 1550K and sodium content of 50g.

In addition, at the low sodium content of 20g and 30g, it can be found that the oxygen content has a significant effect on the detitanation of molten iron, as well as on the dephosphorization. Fig. 5 shows the effect of oxygen content on the detitanation and dephosphorization from molten iron at the temperature of 1550K with sodium content of 20g and 30g, respectively. As can be seen from Fig. 5, with the increase of oxygen content, titanium in molten iron rapidly drops to an extremely low value at first, and then the phosphorus content begins to decline. This indicates that the oxygen entering the molten iron first reacts with the titanium, and then the remaining oxygen reacts with the phosphorus. The removal of titanium from molten iron precedes the removal of phosphorus. This is because the oxidation potential of phosphorus in molten iron is higher than that of titanium [28~29].

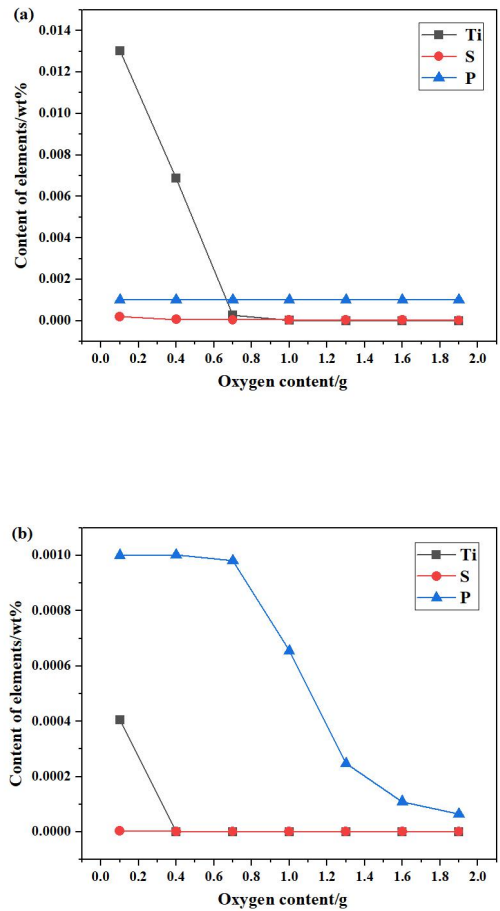

**Fig. 5.** Effect of oxygen content on the detitanation and dephosphorization from molten iron at the temperature of 1550K with sodium content of 20g and 30g, respectively.

*Effect of sodium content on titanium removal*

At the temperature of 1550K and oxygen content of 0.8g, the addition amount of sodium is changed from 20g to 60g, so as to explore the influence of the addition amount of sodium on the removal effect of titanium from molten iron. Fig. 6 shows the effect of sodium content on titanium removal from molten iron. As can be seen from Fig. 6, with the increase of sodium content from 20g to 60g in molten iron, the titanium content in molten iron gradually increases. The results show that too much sodium additives hinder the removal of titanium from molten iron and cause the increase of titanium content in molten iron. Thus, from a thermodynamic point of view, when sodium is added too much, the content of titanium increases, which hinders the removal of titanium from molten iron. This is probably due to the excess sodium content in the molten iron, which consumes oxygen and reduces the oxidation of titanium. However, in the actual production, the volatilization of $Na_2CO_3$ under high temperature should be considered, and the amount of $Na_2CO_3$ added should be appropriate.

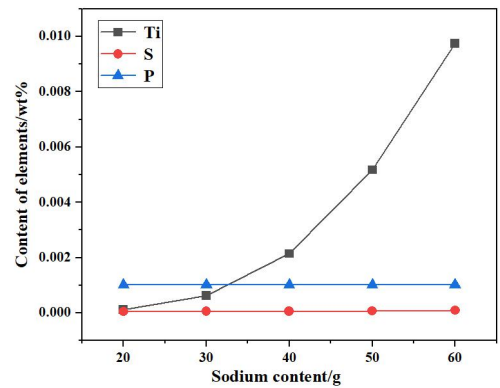

**Fig. 6.** Effect of sodium content on titanium removal from molten iron at the temperature of 1550K and oxygen content of 0.8g.

**Conclusions**

(a) In the process of large amounts of sodium treatment of molten iron, temperature has a great effect on the removal of titanium from molten iron. There is a critical temperature for titanium removal from molten iron. The temperature of titanium removal from molten iron needs to be below the critical detitanium temperature. Moreover, thermodynamic calculation shows that relatively low temperature is favorable for titanium removal from molten iron.

(b) The oxygen content in molten iron affects the critical temperature of titanium removal. With the increase of oxygen content in molten iron, the critical detitanium temperature increases. Moreover, with the increase of oxygen content in molten iron, the removal effect of titanium from molten iron is significantly enhanced.

(c) The oxygen reacts first with titanium in molten iron, and then the remaining oxygen reacts with phosphorus. The removal of titanium from molten iron is earlier than that of phosphorus. High oxygen content in molten iron is conducive to the removal of titanium and phosphorus together.

(d) The addition amount of $Na_2CO_3$ in the smelting system should be moderate considering the thermodynamic equilibrium calculation and the loss of sodium carbonate at high temperature.

**Acknowledgment**

The authors gratefully acknowledge financial support for this work from the National Key R&D Program of China (2018YFC1900500), Supported by the Strategic Priority Research Program of the Chinese Academy of Sciences (XDC04010100), Key Research Program of Frontier Sciences of Chinese Academy of Sciences (QYZDJ-SSW-JSC021), National Natural Science Foundation of China (21908231), Special Project for Transformation of Major Technological Achievements in HeBei province (19044012Z), and Province Key R&D Program of Hebei (20374105D).